\documentclass{webofc}

\usepackage[varg]{txfonts}   
\usepackage{hyperref}
\usepackage{url}
\usepackage{caption}
\usepackage{subcaption}
\RequirePackage{orcidlink}
\hypersetup{colorlinks=true,citecolor=blue,urlcolor=blue,linkcolor=blue}
\begin{document}
\title{Full Event Interpretation at Belle II: Current Status and Recent Developments}
%
%

\author{\firstname{Rahul} \lastname{Tiwary}\inst{1}\orcidlink{0000-0002-5887-1883}\fnsep\thanks{\email{rahul.tiwary@tyl.in2p3.fr}} \and
        \firstname{Mattia} \lastname{Marfoli}\inst{2}\orcidlink{0009-0008-5596-5818} \and
        \firstname{Vidya Sagar} \lastname{Vobbilisetti}\inst{2}\orcidlink{0000-0002-4399-5082} \and
        \firstname{Karim} \lastname{Trabelsi}\inst{1, 2}\orcidlink{0000-0001-6567-3036}
}

\institute{Toshiko Yuasa Laboratory (TYL), Tsukuba
\and
           Centre de Physique des Particules de Marseille (CPPM), Marseille
          }

\abstract{The Full Event Interpretation (FEI) algorithm is a central component of the Belle II analysis framework, designed for the efficient reconstruction of exclusive $B$-meson decays. The FEI performs a hierarchical reconstruction of hadronic and semileptonic final states, using multivariate classification techniques to tag one of the two $B$ mesons produced in electron–positron collisions. The FEI is particularly important for studies of signal-side decays involving missing energy. By reconstructing one $B$ meson with the FEI, the kinematics of the other, signal-side $B$ meson are tightly constrained, leading to improved background rejection.
In this contribution, we present the design, implementation, and recent developments of the FEI. We outline the reconstruction strategy, candidate selection, and multistage multivariate training based on gradient-boosted decision trees. Recent improvements to the FEI workflow, including improvements in the simulated data used for training, reconstruction of intermediate particles, and improved robustness of the multivariate inputs, are discussed in detail.
Beyond algorithmic updates, we also discuss newly developed techniques for the calibration and performance evaluation of the FEI. These include refined procedures to calibrate the efficiency, check the performance across decay channels and data-taking conditions, and application of reweighting techniques to diagnose and mitigate discrepancies between simulation and data. 
Finally, we briefly discuss ongoing and future developments aimed at further improving reconstruction performance and computational efficiency.}
\maketitle
\section{Introduction}
\label{intro}
The Belle II~\cite{belle2tdr} experiment operates at the SuperKEKB~\cite{superkekb} asymmetric-energy electron–positron collider in Tsukuba, Japan. By colliding electrons and positrons at a center-of-mass energy corresponding to the $\Upsilon(4\text{S})$ resonance, pairs of $B$ mesons are produced in a clean experimental environment. The hermetic nature of the Belle II detector and the well-defined initial state provide unique opportunities for studying physics of $B$ mesons. 

A common method to study a $B$ decay ($B_{\mathrm{sig}}$) is to take a set of tracks and clusters from the event and reconstruct the corresponding final state. Such a prescription, although pretty useful in most cases, is not sufficient for all $B$ decays. A significant fraction of the Belle II physics programme relies on final states containing neutrinos or invisible particles. Examples include measurements of semileptonic decays used to determine the CKM matrix elements $V_{cb}$ and $V_{ub}$, tests of lepton flavour universality such as $R(D)$ and $R(D^*)$, and searches for rare or forbidden processes involving $\tau$ leptons or invisible particles. While studying the missing energy final states, reconstructing only the signal-side $B$ decay alone is often insufficient to fully constrain the event kinematics. To address this challenge at Belle II, the companion $B$ meson in the event, commonly referred to as the tag-side $B$ meson ($B_{\mathrm{tag}}$) is reconstructed with dedicated algorithms. The procedure to reconstruct the tag-side $B$ meson is in general referred to as \textbf{tagging}. Once the kinematics of the tag-side is known, further constraints can be imposed on the signal-side $B$ meson, enabling the reconstruction of missing kinematic quantities and also improving background suppression.

\section{$B$-tagging at Belle II}
\label{sec-1}
There are two principal tagging approaches used in Belle II analyses, \textit{Exclusive tagging} and \textit{Inclusive tagging}. In case of exclusive tagging, the $B_{\mathrm{tag}}$ is reconstructed in a set of well known final states. Instead, for inclusive tagging, all the remaining tracks and clusters of the event after reconstruction of $B_{\mathrm{sig}}$ are attributed to the $B_{\mathrm{tag}}$ in order to infer the kinematics of the event.  

The performance of a tagging approach is typically characterized by two primary quantities:
\begin{itemize}
    \item \textbf{Tagging efficiency}: The efficiency to reconstruct the $B_{\mathrm{tag}}$. The tagging efficiency can be expressed as $\epsilon_{\text{Tag}} = \sum^{N}_{i=1} \epsilon_{i}\times\mathcal{B}_{i}$, where $N$ is the total number of $B$ decays reconstructed in the tag-side, and $\epsilon_{i}$, $\mathcal{B}_{i}$ are the reconstruction efficiency and branching fraction for the $i^{\text{th}}$ tag-side $B$ decay, respectively. 
    \item \textbf{Purity}: The fraction of correctly tagged $B$ candidates among all the candidates reconstructed by the tagging procedure.
\end{itemize}

The tagging efficiency of inclusive tagging is $\approx \mathcal{O}(100\%)$, since we are essentially reconstructing all possible $B$-tag candidates. Nevertheless, an high tagging efficiency comes with a relatively low purity due to lack of strong constraints on the $B_{\mathrm{tag}}$. On the other hand, in case of exclusive tagging, we reconstruct the $B_{\mathrm{tag}}$ via an handful of exclusive final states. Depending on the set of final states, the exclusive tagging can be further classified into two categories, \textit{hadronic tagging} and \textit{semileptonic tagging}. The tagging efficiency of hadronic tag is $\approx \mathcal{O}(0.1\%)$, and that of semileptonic tag is $\approx \mathcal{O}(1\%)$, where the difference can be largely attributed to the higher branching fraction of semileptonic $B$ decays compared to the hadronic $B$ decays. Despite the lower tagging-efficiency, the hadronic tagging is still useful since it provides a higher purity sample compared to semileptonic tag, and offers stronger constraints on the event by fully reconstructing the tag-side.

The reconstruction is performed in multiple stages. In the first stage, final-state particles (FSPs) are reconstructed from charged tracks ($e$, $\mu$, $K$, $\pi$, $p$) and neutral clusters ($\gamma$). These FSPs are then combined to reconstruct secondary particles such as $K^{0}_{S}$, $\Lambda$, $\pi^{0}$. In the subsequent stages, higher-level intermediate particles, including $D$, $D^{*}$, and $\Lambda_{c}$ mesons, are reconstructed using different combinations of final-state and secondary particles. Finally, $B$-meson candidates are formed by combining the reconstructed intermediate particles and, where appropriate, final-state particles.

At every stage of the reconstruction chain, multivariate classifiers based on boosted decision trees (BDTs)~\cite{FastBDT} are employed to distinguish correctly reconstructed candidates from backgrounds. Separate classifiers are trained for different particle species and decay modes. The input variables include particle identification variables, kinematic quantities, vertex information, kinematics of the decay products, and BDT outputs of the decay products. This hierarchical strategy allows information from all stages of the decay chain to be propagated to higher-level candidates. The resulting classifier output provides a powerful estimate of the signal probability for each reconstructed $B-$tag candidate.

\section{Calibration of FEI}
\label{sec-2}
The FEI relies heavily on simulated samples (MC) for training the BDTs. Hence, the performance must be carefully calibrated using collision data. Imperfections in detector simulation, decay modelling, and reconstruction efficiencies can lead to discrepancies between data and MC. 

In order to measure these quantities, we employ well-known $B$ decays, also referred to as control channels, and pair them up with FEI-tagged $B$ candidates. A classic method is to use the inclusive semileptonic decay ($ B \rightarrow X \ell \nu $) as the control channel~\cite{Sutcliffe:2021xG}. By comparing the observed yields of FEI tagged control channel in data and MC, one can extract correction factors for the FEI. These calibration factors encompass the data-MC differences associated with the tagging procedure, ensuring reliable physics measurements.

A major recent development has been the introduction of a new calibration strategy based on the decay channel 
$B \rightarrow D^{(*)}\pi$~\cite{vobbilisetti:tel-04306669} for the hadronic FEI. In this method, a charged pion is paired with a FEI tag candidate reconstructed in hadronic modes. The system recoiling against the charged pion and hadronic FEI pair is examined for the presence of a $D$ or $D^{*}$ meson. The resulting control sample is statistically potent and, importantly, orthogonal to the previously used semileptonic calibration channels. The calibration factors are determined through the ratio: $C_{\mathrm{FEI}} = N(D\pi)_{\mathrm{Data}}/N(D\pi)_{\mathrm{MC}}$, where the yields are extracted from fits to the recoil mass. Another advantage of the $B \rightarrow D^{(*)}\pi$ calibration channel is the applicability of the $_s\mathcal{P}lot$~\cite{sPlot} technique for background subtraction. This enables detailed studies of the kinematic properties of the $B-$tag candidates and provides valuable insight into residual discrepancies between data and MC. The introduction of this calibration mode has significantly improved the understanding of FEI performance and has become an important validation tool for ongoing developments.

\section{Improvements to FEI}
Aided with the $B \rightarrow D^{(*)}\pi$ mode, several improvements to the hadronic FEI algorithm were performed. In this section, we discuss each one of them.

\subsection{Simulation model}
The performance of the FEI is intrinsically linked to the quality of the simulated samples used during training. Any discrepancy between simulated and real $B$-meson decays can directly affect reconstruction efficiency and classification performance. A detailed investigation of hadronic $B$-meson decay modelling revealed several deficiencies in the simulation. Many high-multiplicity decay modes are poorly measured experimentally and therefore rely on phenomenological models. In some cases, inclusive branching fractions had been interpreted as exclusive non-resonant contributions, potentially leading to double counting.

\begin{figure}[htbp]
    \centering
    \begin{subfigure}[b]{0.45\textwidth}
        \includegraphics[width=\textwidth]{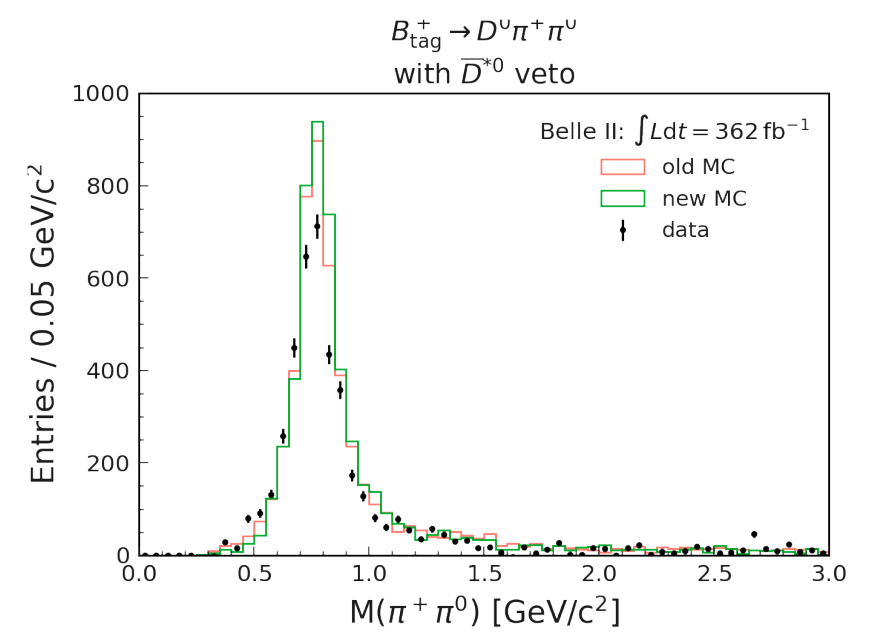}
        \caption{Invariant mass of two pions for the decay $B^- \rightarrow D^0 \pi^-\pi^0$.}
        \label{fig:Dpipi0}
    \end{subfigure}
    \begin{subfigure}[b]{0.45\textwidth}
        \includegraphics[width=\textwidth]{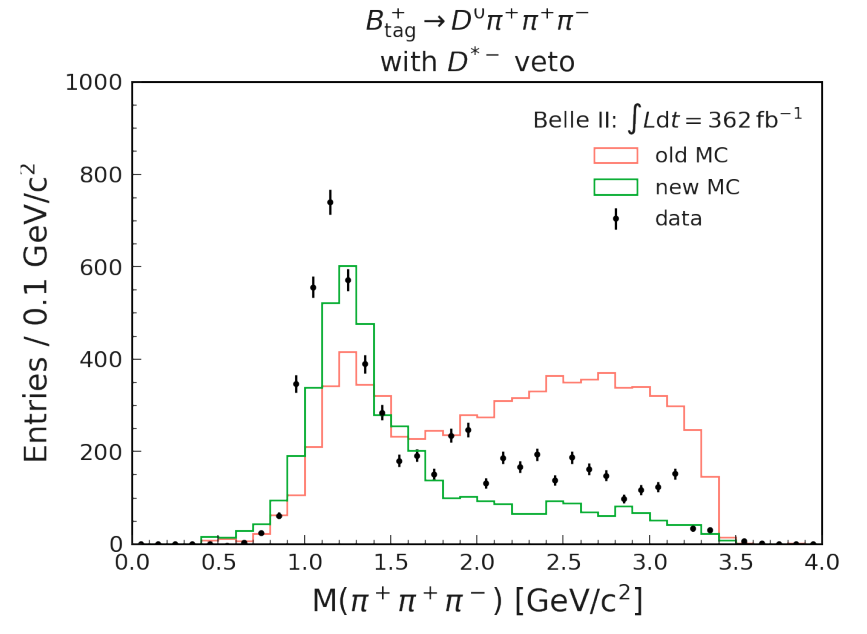}
        \caption{Invariant mass of three pions for the decay $B^- \rightarrow D^0 \pi^-\pi^+\pi^-$.}
        \label{fig:D3pi}
    \end{subfigure}
    \caption{Comparison of invariant mass between data, the old, MC and the new MC.}
    \label{fig:Invariant_mass_comparison}
\end{figure}

An important insight of this study was the fact that a number of hadronic $B$ decay modes suffer from large statistical uncertainty in the branching fraction measurement. One such example is the decay $B^- \rightarrow D^0 \pi^-\pi^0$, where the dominant contribution is from the  $B^- \rightarrow D^0 \rho^-$ process. The current MC uses the branching fraction value $(1.34\pm0.18\%)$ for the $B^- \rightarrow D^0 \rho^-$ component, which is based on the CLEO measurement~\cite{PhysRevD.50.4368}. The invariant mass of two pions for $B^- \rightarrow D^0 \pi^-\pi^0$ mode after vetoing $D^{*0}$ contribution is shown in Fig.~\ref{fig:Dpipi0}. We observe that the MC overestimates the $\rho$ component, which could be an artifact of small statistics (0.9 $\rm{fb}^{-1}$) used by $B^- \rightarrow D^0 \rho^-$ measurement. A recent measurement by Belle~II~\cite{Phys.Rev.D109.2024} reported the branching fraction to be $(0.94\pm0.05\%)$, a factor of two improvement in the precision with respect to the CLEO measurement, and a slightly reduced central value. The updated branching fraction of $B^- \rightarrow D^0 \rho^-$ mode will be used for the future MC samples. This example illustrates the importance of revisiting the measurements for such hadronic $B$ decays to improve the MC modelling, which can have significant impact on background modelling for precision measurements of rare $B$ decays, and can potentially improve the FEI performance.

Another illustrative example is the decay $B^- \rightarrow D^0 \pi^-\pi^+\pi^-$ measured by LHCb~\cite{PhysRevD.84.092001} which only reports the branching fraction of the process. No further information is provided on the contributing processes with intermediate resonances. Studies of the three-pion invariant mass distribution indicated that the decay is dominated by intermediate resonant contributions, particularly $B^- \rightarrow D^0 a_1^-$, rather than purely non-resonant production.  Updating the simulation to reflect this resonance structure substantially improves the agreement between data and MC samples, as shown in Fig.~\ref{fig:D3pi}. 

In summary, the MC sample was updated to accommodate the latest measurements of branching fractions for $B$ decays, the contribution of intermediate resonances were modified to align the MC with data, and double counting of non-resonant contributions were removed. These modifications resulted in significantly improved data–simulation agreement across a wide range of observables. Most importantly, the overall FEI calibration factor improved dramatically, increasing from approximately 0.65 to 0.83, accompanied by an improvement in tagging purity from 0.56 to 0.63. 

\subsection{Reconstruction of neutrals}
In addition to improvements in decay modelling, significant gains have been achieved through refinements of photon and $\pi^0$ reconstruction. Historically, the selection criteria applied to photons in the Belle II FEI were more restrictive than those used in the predecessor Belle experiment\cite{FR}. These tighter requirements reduced the reconstruction efficiency. To recover efficiency, the photon selections were revised and made consistent with those used in Belle's tagging framework. Furthermore, a mass-constrained fit is now applied to reconstructed $\pi^0$ candidates. This improves the energy and momentum resolution of the neutral pions by enforcing the known $\pi^0$ mass during reconstruction.

These modifications positively impacted both the hadronic and semileptonic FEI channels. Studies performed using updated training samples demonstrated substantial increases in reconstructed signal yields. These gains illustrate how detector-level reconstruction improvements can propagate through the FEI hierarchy and enhance overall tagging performance.

\section{FEI performance}
Following the implementation of the simulation and reconstruction improvements described above, a new FEI training campaign was carried out using updated MC samples and modified algorithm for reconstruction of neutrals. The performance of the retrained FEI was evaluated using the $B \rightarrow D^{(*)}\pi$ calibration channel. The resulting calibration factors for the hadronic FEI were found to be close to unity:

$$
C(B^0) = 1.06 \pm 0.02 \pm 0.04, \;
C(B^+) = 1.22 \pm 0.04 \pm 0.05.
$$

\begin{figure}[htbp]
    \centering

    \begin{subfigure}[b]{0.475\textwidth}
        \includegraphics[width=\textwidth]{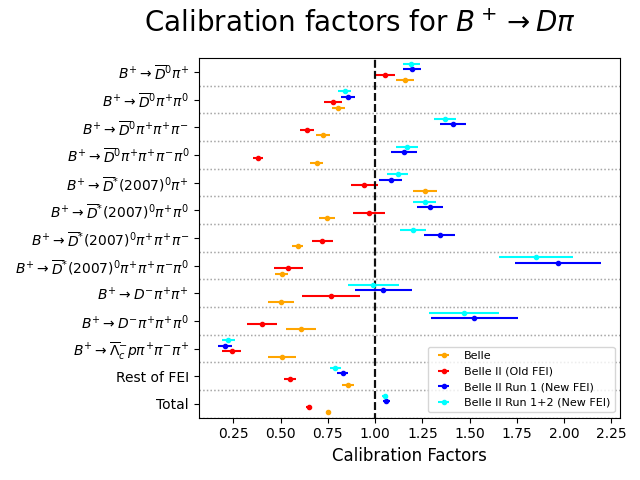}
        \caption{Calibration factors for hadronic FEI.}
        \label{fig:fei_cf}
    \end{subfigure}
    \begin{subfigure}[b]{0.475\textwidth}
        \includegraphics[width=\textwidth]{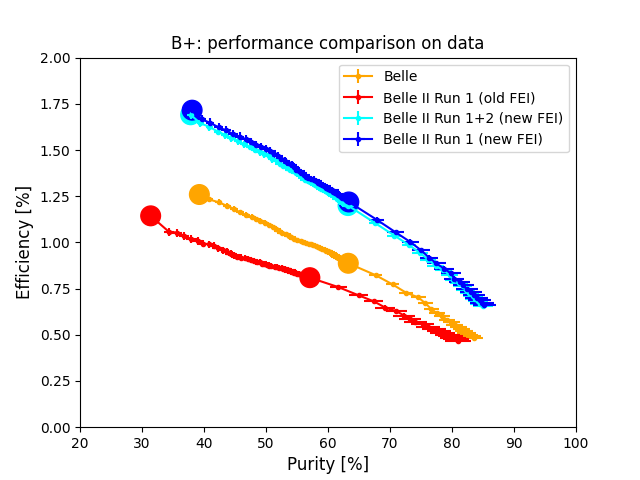}
        \caption{Performance comparison of hadronic FEI.}
        \label{fig:fei_eff_pur}
    \end{subfigure}

    \caption{Summary of FEI performance.}
    \label{fig:FEI_Performance}
\end{figure}

A mode-by-mode breakdown of the calibration factors and the comparison of performance of hadronic FEI between different FEI versions is illustrated in Fig.~\ref{fig:FEI_Performance}.
These results indicate a substantially improved agreement between data and simulation compared with earlier FEI versions. Perhaps more importantly, the updated FEI delivers a significant increase in tagging efficiency. Relative to the previous Belle II implementation, improvements of approximately 50\% for $B^+$ tagging, and 30\% for $B^0$ tagging, have been observed.

\section{Future prospects}
Although considerable progress has been achieved, several opportunities remain for further improving FEI performance.

\subsection{Improved Decay Modelling}

As discussed earlier, many hadronic $B$-meson decay modes remain poorly measured. Existing simulations often rely on simplified phase-space descriptions despite the presence of rich resonance structures. The MC mismodeling results in the calibration factor of such modes being far away from unity. A good example is the mode $B^{+}\to\overline{\Lambda_{c}}p\pi^{-}\pi^{+}$ measured by CLEO~\cite{PhysRevD.66.091101}. Modelled with an assumption of uniform phase-space in the MC, the channel exhibits rich topology of intermediate resonances. Improving the measurements of branching fractions as well as characterizing the relative contribution of resonances will result in realistic MC modelling, and consequently more accurate FEI training.

\subsection{Revisiting Input Features}

The current hadronic FEI training includes the variable $\Delta\text{E}$, which is the difference between the energy of reconstructed $B$-tag candidate and half the beam energy. The variable $\Delta\text{E}$ provides powerful discrimination between signal and background candidates. The current hadronic FEI doesn't have a robust variable which can be used as fit variable to estimate the number of tagged $B$ candidates. Hence, removing $\Delta\text{E}$ from the hadronic FEI training can allow it to be used later as an independent fit observable in physics analyses. Furthermore, one can also experiment with improved particle identification variables~\cite{PIDNN} as input features to the FEI algorithm.

\subsection{Improved multivariate methods}

The rapid development of modern machine-learning techniques offers exciting prospects for future tagging algorithms. One promising direction is GraFEI~\cite{GraFEI}, a graph-neural-network-based reconstruction framework that represents decay chains as graphs rather than fixed hierarchical structures. Such approaches may better capture correlations among decay products and provide improved performance relative to traditional BDTs.

\section{Summary and conclusion}

The Full Event Interpretation is a cornerstone of the Belle II physics programme and provides the experiment's most powerful exclusive $B$-tagging capability. By reconstructing hundreds of decay channels through a hierarchical multivariate framework, the FEI enables precision measurements and searches involving final states with missing energy.

Recent efforts have focused on improving the agreement between data and simulation through dedicated calibration studies, enhanced decay modelling, and optimized reconstruction of neutral particles. The introduction of the $B \rightarrow D^{(*)}\pi$ control channel has provided a powerful new calibration tool, facilitating improvements to MC modelling to get substantially improved calibration factors and tagging purity.

A retrained FEI based on these developments delivers significantly improved performance, with tagging-efficiency gains of up to 50\% relative to earlier Belle II implementations. Future improvements are expected from better decay modelling, refined input features, and the adoption of modern machine-learning techniques. These developments ensure that the FEI will continue to play a central role in maximizing the physics reach of Belle II throughout the coming years.

\bibliography{bibliography} 

\end{document}